\documentclass[aps,prx,notitlepage,twocolumn,superscriptaddress,nofootinbib]{revtex4-2}
\usepackage{siunitx}
\usepackage{amsmath}
\usepackage{amssymb}
\usepackage{bbold}
\usepackage{color}
\usepackage{xcolor}
\usepackage{comment}
\usepackage{tikz}
\usepackage{pgfplots}
\pgfplotsset{compat=1.5}
\usepackage{graphicx}
\usepackage[colorlinks=true, linkcolor=blue, citecolor=blue, pdfencoding=auto]{hyperref}
\usepackage{physics}
\usepackage[utf8]{inputenc}
\usepackage{comment}
\usepackage{ulem}

\newcommand{\br}{\mathbf{r}}
\newcommand{\bk}{\mathbf{k}}

\newcommand{\bR}{\mathbf{R}}

\newcommand{\bG}{\mathbf{G}}

\def\phdag{{\phantom\dagger}}

\newcommand{\moire}{moir\'e}
\newcommand{\Moire}{Moir\'e}

\definecolor{todogray}{gray}{0.5}

\begin{document}

\title{Cell Natural Orbitals in Quantum Materials}

\author{Harshitra Mahalingam}
\affiliation{Department of Physics, Columbia University, New York, NY 10027, USA}

\author{Nishchhal Verma}
\affiliation{Department of Physics, Columbia University, New York, NY 10027, USA}

\author{Daniel Mu\~{n}oz-Segovia}
\affiliation{Department of Physics, Columbia University, New York, NY 10027, USA}

\author{Raquel Queiroz}
\email{raquel.queiroz@columbia.edu}
\affiliation{Department of Physics, Columbia University, New York, NY 10027, USA}
\affiliation{Center for Computational Quantum Physics, Flatiron Institute, New York, NY 10010, USA}

\begin{abstract}
Understanding correlated quantum matter starts with an accurate model of the single-particle states that interact at low energies: their dispersion, band geometry, orbital content and charge density. 
In many cases, notably the topological bands of {\moire} materials, it is not straightforward to find a real-space description with a few local orbitals that accomplishes this task.
Here we provide a systematic way to identify the local degrees of freedom that best capture the band geometry and charge density of any chosen set of bands. We use the unit-cell one-particle reduced density matrix (UC-1pRDM), obtained by restricting the projector onto the target bands to a single unit cell. Its eigenstates, which we call cell natural orbitals (CNOs), form a local, symmetric basis uniquely determined by the Bloch wavefunctions and the choice of real-space partition. Their eigenvalues measure the occupation of each CNO in the target bands, quantifying entanglement across unit-cell boundaries and the importance of multi-orbital character. 
A set of CNOs that maximizes total spectral weight and reproduces the target band symmetries provides optimal trial states for Wannierization. We exemplify this by constructing a lattice model for twisted bilayer WSe$_2$ that tracks the orbital content across twist angles.
\end{abstract}

\maketitle

\section{Introduction}
Physical intuition of interacting quantum materials is often built out of a minimal description of the relevant low-energy degrees of freedom.
While the understanding of wavefunctions in crystalline solids, like their quantum geometric properties~\cite{Torma2023_prl_commentary,Yu2025QuantumMaterials,Rossi2021}, usually relies on momentum space, a real-space model provides a useful basis for correlated systems thanks to the locality of electronic interactions.
To accurately capture the physics of the system, a model should reproduce the electronic dispersion, symmetry and quantum geometric properties of the wavefunctions. 
Identifying a minimal set of local orbitals that fulfills these criteria, however, is a nontrivial task.

Recent advances in {\moire} materials \cite{Andrei2021_rev, Nuckolls2024AMaterials} and growing interest in flat-band physics \cite{Aoki2025FlatSuperconductivity} have emphasized that correlation phenomena depend not only on the band energies but also on the momentum dependence of the Bloch wavefunctions themselves.
This wavefunction information resides in how the cell-periodic Bloch states $|u_{\bk n}\rangle$ evolve across the Brillouin zone and is known as quantum geometry \cite{Provost1980RiemannianStates, Gao2025QuantumSystems}. 
Because the cell-periodic wavefunctions describe the charge distribution within a unit cell \cite{Vanderbilt2018Berry}, preserving quantum geometry amounts to faithfully reproducing the intracell charge density \cite{Ingham2026Real-SpaceZeros} and the associated dipole matrix elements \cite{Verma2025InstantaneousInsulators}.
Although quantum geometry is ubiquitous in materials, it becomes physically important when the orbital composition of the low-energy states varies strongly across momentum space \cite{Verma2026_nrr}. 
In such systems, 
quantum geometry plays a central role in determining the effective interactions \cite{MoralesDuran_nonlocal_2022, Wu_firstTMD} and low-energy physics \cite{Crepel2024_prr, Qiu_Fengcheng_PRX, Zang_Millis_TMD_2021}, making its faithful representation an essential requirement for the effective real-space local model.

The central problem is therefore to determine the appropriate real-space orbitals corresponding to a set of target bands. 
Most existing approaches use a set of trial
orbitals and then optimize their overlap with the target bands \cite{Cloizeaux1964AnalyticalFunctions}. Maximally localized Wannier functions yield localized orbitals \cite{MarzariMLWFRMP2012}, while disentanglement methods extend this construction to entangled band manifolds \cite{SouzaMarzariVanderbilt2001}. 
Automated approaches such as SCDM remove the need for physically motivated initial projections \cite{Damle2015,Damle2017, Vitale2020AutomatedWannierisation}.
However, these methods still require specifying the number of Wannier orbitals, and therefore do not optimize the minimal local Hilbert space required by the bands itself. 
\begin{figure*}
    \centering
    \includegraphics[width=2\columnwidth]{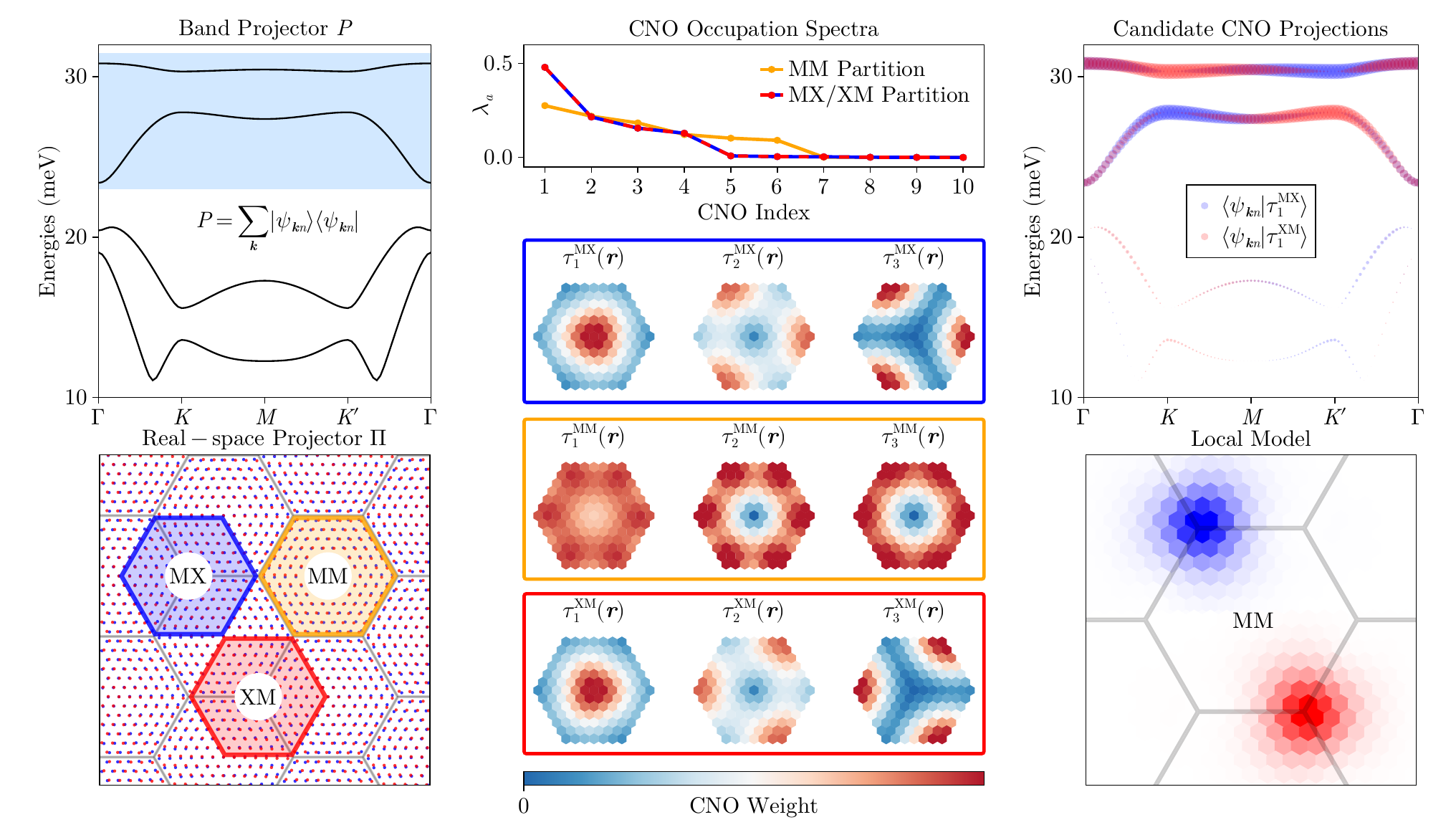}
    \caption{
    Schematic overview of the cell natural orbital construction.
    Starting from a target band projector $P$, the projector is restricted to a single real-space unit cell through the local projector $\Pi$ to form the unit-cell one-particle reduced density matrix $L=\Pi P\Pi$.
    Diagonalization yields the cell natural orbitals (CNOs) $|\tau_a\rangle$ and their occupations $\lambda_a$.
    The CNOs provide compact local orbitals labeled by their symmetry and Wyckoff position, while the occupations rank their importance within the target bands. 
    Candidate tight-binding models are constructed by selecting CNO subsets, testing their ability to span the target bands through the overlap matrix $\mathcal{S}_{\bk}$, and optimizing their total spectral weight.
    A chosen CNO subset is then used as trial states in a single-shot Wannierization procedure to obtain a local model that preserves the quantum geometry of the target bands.
    }
    \label{fig:CNO_schematic}
\end{figure*}

A complementary route is offered by topological quantum chemistry~\cite{Bradlyn2017,Cano2018}, which classifies bands according to elementary band representations (EBRs) induced from localized orbitals at specific Wyckoff positions. 
While it identifies symmetry-compatible atomic limits \cite{Zak1982}, it does not identify which local orbitals have the most overlap with the wavefunctions over the target bands.

In this work, we introduce a systematic construction of local orbitals using only the band projector as input.
The central object is the unit-cell one-particle reduced density matrix (UC-1pRDM), $L=\Pi P\Pi$, where $P$ projects onto the target Bloch bands and $\Pi$ restricts the Hilbert space to a single unit cell.
The eigenstates of this operator, which we call \textit{Cell Natural Orbitals} (CNOs), provide the local orbital basis intrinsically selected by the target Bloch bands (see Fig.~\ref{fig:CNO_schematic}).
Although natural orbitals have been known for decades \cite{Lowdin1955}, their periodic counterpart in the non-interacting limit, CNOs, have remained unexplored.

The eigenvalues $\lambda_a$ provide a ranking of the local degrees of freedom. 
They quantify the weight of each CNO inside the target bands and are simultaneously the singular values of the map between local orbitals and the band subspace, as well as the eigenvalues of the real-space correlation
matrix governing unit-cell entanglement.
Dominant occupation of a single CNO indicates that the set of bands is well described by a single orbital, while a fractional occupation spectrum reveals the necessity of multiple orbitals.

CNOs provides a connection between local orbital structure and topology. 
For Chern bands, the projected CNO envelope functions necessarily contain zeros enforced by the topology. 
These irremovable zeros prevent any single compact orbital from spanning the band manifold and provide a real-space manifestation of topological obstruction. 
Other than topology, zeros can arise from incompatible Wyckoff positions in obstructed atomic insulators.
CNOs inherit symmetry properties from symmetry-preserving unit-cell partitions and therefore identify candidate EBRs.
The completeness of the basis depends on the momentum-space overlap matrix between CNOs and target bands.

The goal of constructing real-space models for topological bands has motivated several recent approaches to overcoming the associated Wannier obstruction. These include relaxing symmetry requirements, like translational symetry~\cite{MonsenClasseenSupercellWannier2024,ColeVanderbiltReducedWannierRepresentation2025,Fischer2024SupercellGraphene} or valley conservation~\cite{Kang2018SymmetryBands,2020textitAbSystems}, enlarging the Hilbert space by introducing auxiliary orbitals outside the target manifold~\cite{Po2019FaithfulGraphene,Carr2019DerivationApproach,Calderon2020InteractionsGraphene}, as explored in heavy-fermion-like models \cite{song2022magic,Calugaru2023TwistedParameters}, and compact molecular orbitals \cite{Xie_cmo_1,Xie_cmo_2,Xie_cmo_3,Xie_cmo_4, Li_cmo_5}.
In this work, we address this challenge by using CNOs as trials for Wannierization, and show a specific application for twisted WSe$_2$.

The remainder of the paper is organized as follows. Sec.~\ref{sec:UC1pRDM} introduces the UC-1pRDM and the CNOs, relates the CNO occupation spectrum to unit-cell entanglement, and establishes its dependence on the choice of unit-cell partition. Sec.~\ref{subsec:constraints_CNO_symmetry_topology} shows that each CNO carries a Wyckoff position and a site-symmetry irrep, and uses the resulting envelope functions to separate the irremovable zeros of Chern bands from the removable zeros enforced by symmetry. Sec.~\ref{sec:constructing_TB} describes the construction of tight-binding models, giving the criteria that decide whether a chosen set of CNOs spans the target bands and how to rank competing sets. Sec.~\ref{sec:tWSe2} extends the construction to continuum models and applies it to twisted bilayer WSe$_2$.

\section{Cell Natural Orbitals}
\label{sec:UC1pRDM}
Consider a crystal with Bravais lattice vectors $\bR$ and $N_\alpha$ internal degrees of freedom per unit cell, labeled by an index $\alpha$ that can include orbital, sublattice, spin and layer quantum numbers. 
The creation operator $c^\dag_{\bR, \alpha}$ places a fermion in unit cell $\bR$ with internal label $\alpha$. 
The single-particle Bloch states are defined by their momentum label $\bk$ in the Brillouin zone and band label $n$, such that
\begin{equation}
    |\psi_{\bk n} \rangle \equiv c^\dagger_{\bk n} | 0 \rangle = \frac{1}{\sqrt{N_k}}
    \sum_{\bR,\alpha}
    e^{i\bk\cdot(\bR+\br_\alpha)}\,
    u_{\alpha,\bk n}\,
    c^\dagger_{\bR\alpha} | 0 \rangle \, .
    \label{eq:bloch_expand}
\end{equation}
where $N_k$ is the number of unit cells, and $u_{\alpha,\bk n}$ are the Bloch amplitudes. 
The $\br_\alpha$ entering the Fourier transform depend on the orbital embedding choice \cite{Haldane2014AttachmentMetals,Lim2015GeometryInterferometry,Herzog-Arbeitman2022SuperfluidBands,Huhtinen2022RevisitingTouchings}; for example, in the physical embedding $\br_\alpha$ is the physical position of orbital $\alpha$ within the unit cell, whereas in the periodic embedding $\br_\alpha=0$. 
We can always collect the amplitudes into a cell-periodic vector $|u_{\bk n}\rangle$ in orbital space. 
We emphasize that the basis states underlying $|u_{\bk n}\rangle$ and $|\psi_{\bk n}\rangle$ are distinct, despite the use of the same ket notation. The cell periodic state $|u_{\bk n}\rangle$ lives in the internal orbital space, whereas the Bloch state $|\psi_{\bk n}\rangle$ lives in the full lattice Hilbert space.

\subsection{Local Orbitals from Target Bands}

The connection between local orbitals and target bands can be formulated in terms of two projection operators. The target manifold is defined by a set of Bloch states ${|\psi_{\bk n}\rangle}$ with restrictions on the band index $n$ and crystal momentum $\bk$. We define the corresponding projector as
\begin{equation}
P =  \sum_{\bk,n \in P} |\psi_{\bk n} \rangle\langle\psi_{\bk n}| \, .
\label{eq:def:P_M}
\end{equation}
The local orbital subspace is spanned by states constrained to a single cell,
\begin{equation}
\Pi_\bR = \sum_{\alpha \in {\rm u.c.}} | \bR,\alpha \rangle \langle \bR,\alpha | ,
\label{eq:projector_PR0}
\end{equation}
which annihilates any state with weight outside cell $\bR$. 
Choosing the reference unit cell $\bR = 0$ and suppressing the $\bR$ label, we define the operator
\begin{equation}
L = \Pi P \Pi .
\label{eqn:L_R_definition}
\end{equation}
The eigenstates of this operator are local orbitals that maximize their overlap with the low-energy subspace.

This operator is the correlation matrix restricted to a single unit-cell partition or equivalently the unit-cell one-particle reduced density matrix (UC-1pRDM) \cite{Peschel2003JPhysA, Peschel2009}. 
For a periodic system, translation symmetry ensures that the matrix elements of $L$ are identical for every unit cell. In the orbital basis, the matrix elements are
\begin{equation}
L_{\alpha\beta}
= \langle c^\dag_{\beta} c^{\phdag}_\alpha \rangle
= \frac{1}{N_k}
\sum_{\bk,n\in P}
e^{i\bk\cdot( \br_\alpha - \br_\beta)}
\langle\alpha|u_{\bk n}\rangle\langle u_{\bk n}|\beta\rangle
\label{eq:ucRDM_matrix_new}
\end{equation}
which can equivalently be written as
\begin{equation}
L = \frac{1}{N_k} \sum_{\bk,n\in P} V_\bk \, |u_{\bk n}\rangle\langle u_{\bk n}| \, V^\dagger_\bk
\label{eq:ucRDM_matrix}
\end{equation}
where $V_\bk = \mathrm{diag}(e^{i\bk\cdot\br_1},\dots,e^{i\bk\cdot\br_{N_\alpha}})$ is the embedding matrix. Including $V_\bk$ in Eq.~\eqref{eq:ucRDM_matrix} is crucial; excluding it would lead to eigenstates that are not exponentially localized. 

The UC-1pRDM is Hermitian and positive semidefinite. 
Its eigenstates define the cell natural orbitals (CNOs) ${|\tau_a\rangle}$ and the corresponding occupations ${\lambda_a}$,
\begin{equation}
L|\tau_a\rangle = \lambda_a|\tau_a\rangle \, .
\label{eq:CNO_eig}
\end{equation}
The occupations rank the local orbital spectral weight within the band manifold with
\begin{equation}
    \lambda_a = \dfrac{1}{N_k} \sum\limits_{\bk, n \in P} |\langle u_{\bk n} | V^\dag_\bk| \tau_a \rangle|^2 \, .
    \label{eq:lambda_specW}
\end{equation}
The completeness of CNO basis enforces the normalization condition
\begin{equation}
\frac{1}{\langle N_{\mathrm{cell}}\rangle} \sum_a\lambda_a =  1 \, ,
\label{eq:sumrule}
\end{equation}
where $\langle N_{\mathrm{cell}}\rangle$ is the total electron number per unit cell in the target bands.

The UC-1pRDM is not a projector in general, with $L^2\neq L$, and therefore the CNO occupations are generically fractional. 
A band manifold whose quantum geometry is fully captured by a single local orbital has $\lambda_1=1$ and $\lambda_{a>1}=0$. 
Deviations from this limit indicate that multiple unit-cell compact orbitals are required to represent the low-energy wavefunctions.

We emphasize that the UC-1pRDM is a real-space object and therefore it is independent of the orbital embedding choice. In particular, the embedding dependence of $V_\bk$ and $|u_{\bk n}\rangle$ in Eq.~\eqref{eq:ucRDM_matrix} cancels out; for example, in the periodic embedding:
\begin{equation}
L= \frac{1}{N_k} \sum\limits_{\bk,n\in P} |u^{\rm per}_{\bk  n}\rangle \langle u^{\rm per}_{\bk n}| \, .
\label{eq:ucRDM_periodic}
\end{equation}

For tight-binding models, the orbital positions $\br_\alpha$ are explicit and the embedding matrix $V_\bk$ is known. 
For continuum models written in a plane-wave basis, there is no a priori notion of orbital positions. We discuss this in detail in Sec.~\ref{subsec:real_space_grid_plane_wave}.

\subsection{Relation to Projected Density Operator}
\label{subsec:SVD}

The UC-1pRDM is connected to the projected unit-cell charge density operator, as shown below. 
We define the rectangular operator
\begin{equation}
\mathcal{U}\equiv \Pi P,
\label{eq:U_def}
\end{equation}
which maps the target Bloch subspace onto the orbital Hilbert space of a single unit cell.
Its two adjoint products are
\begin{equation}
L=\mathcal{U}\mathcal{U}^\dagger,
\qquad
\Lambda=\mathcal{U}^\dagger\mathcal{U},
\label{eq:two_products}
\end{equation}
where 
\begin{equation}
    \Lambda=P\Pi P 
\end{equation}
is the projected density operator associated with the reference unit cell.
In the Bloch basis,
\begin{equation}
\langle\psi_{\bk n}|\Lambda|\psi_{\bk' n'}\rangle
= \frac{1}{N_k}
\langle u_{\bk n}|V_\bk^\dagger V_{\bk'} |u_{\bk' n'}\rangle =
\frac{1}{N_k}
\langle u^\mathrm{per}_{\bk n}|u^\mathrm{per}_{\bk' n'}\rangle,
\label{eq:Lambda_elements}
\end{equation}
so the matrix elements of $\Lambda$ are the quantum-geometric form factors of the selected bands.

Since $L$ and $\Lambda$ are the two adjoint products of the same rectangular operator, they possess the same nonzero eigenvalues.
Writing the singular value decomposition of $\mathcal{U}$,
\begin{equation}
\mathcal{U}|s_a\rangle
=
\sqrt{\lambda_a}\,|\tau_a\rangle,
\qquad
\mathcal{U}^\dagger|\tau_a\rangle
=
\sqrt{\lambda_a}\,|s_a\rangle,
\label{eq:SVD}
\end{equation}
the left singular vectors $|\tau_a\rangle$ are precisely the CNOs introduced in Eq.~\eqref{eq:CNO_eig}, while the right singular vectors $|s_a\rangle$ are exponentially localized wavefunctions centered at the reference unit cell, and live within the target bands subspace.

We define the corresponding envelope functions by the rescaled Bloch coefficients
\begin{equation}
s_{a,n,\bk}
\equiv
\sqrt{N_k}\,
\langle\psi_{\bk n}|s_a\rangle
=
\frac{1}{\sqrt{\lambda_a}}
\langle u^\mathrm{per}_{\bk n}|\tau_a\rangle,
\label{eq:envelope}
\end{equation}
which, for a given unit cell, satisfy the orthogonality condition
\begin{equation}
 \langle s_a|s_b\rangle = \frac{1}{N_k} \sum_{\bk,n \in P} s^*_{a,n,\bk} s_{b,n,\bk}
=
\delta_{ab}.
\label{eqn:sk_orthonormality}
\end{equation}
For a single isolated band we drop the band index and write $s_{a,\bk}$.
The projected image of a CNO follows directly from Eq.~\eqref{eq:SVD} with
$P|\tau_a\rangle
=
\sqrt{\lambda_a}\,
|s_a\rangle$.

The projected density operator admits the spectral decomposition
\begin{equation}
    \Lambda_{\bk n,\bk' n'} \equiv \langle\psi_{\bk n}|\Lambda|\psi_{\bk' n'}\rangle = \frac{1}{N_k} \sum_a \lambda_a\, s^*_{a,n,\bk} s_{a,n',\bk'},
\label{eq:Lambda_SVD}
\end{equation}
which expresses every quantum-geometric form factor as a sum over orthogonal CNO channels weighted by their occupations.
Retaining a subset of channels therefore gives a low-rank approximation to the form factor matrix.
These channels can also be used to simplify band projected interactions \cite{verma2025local}.

\subsection{Unit-Cell Entanglement and Choice of Real-Space Partition}
\label{subsec:entanglement}

Consider the Slater determinant formed by occupying all Bloch states in the target manifold,
\begin{equation}
|\Psi\rangle
=
\prod_{\bk,n\in P}
c^\dagger_{\bk n}|0\rangle.
\end{equation}
Once we partition the lattice into the reference unit cell and its complement, the reduced density matrix of the unit cell is obtained by tracing over all degrees of freedom outside the partition,
\begin{equation}
\rho_{\rm cell}
=
\mathrm{Tr}_{\overline{\rm cell}}
\left(
|\Psi\rangle\langle\Psi|
\right).
\label{eq:rho_cell}
\end{equation}

Since $|\Psi\rangle$ is a Gaussian state, Wick's theorem implies that $\rho_{\rm cell}$ is completely determined by the one-particle correlation matrix restricted to the unit cell \cite{Peschel2003JPhysA, Peschel2009} given by $L$ in Eq.~\eqref{eq:two_products}. Specifically, one can write
\begin{equation}
    \rho_{\rm cell} = \bigotimes_a \lambda_a |1_a \rangle \langle 1_a | + (1- \lambda_a) |0_a\rangle \langle 0_a |
\end{equation}
where $1$ and $0$ refer to the occupied and unoccupied sector of CNO $a$.
Since the reduced density matrix factorizes over the entanglement modes,
the von Neumann entropy is
\begin{align}
S_{\rm cell}
&=
-\mathrm{Tr}
\left(
\rho_{\rm cell}
\log\rho_{\rm cell}
\right) \nonumber \\
&=
-
\sum_a
\left[
\lambda_a\log\lambda_a
+
(1-\lambda_a)\log(1-\lambda_a)
\right].
\label{eq:Scell}
\end{align}
Modes with $\lambda_a=0$ or $1$ are pure and do not contribute to the entropy,
while $\lambda_a=1/2$ gives the maximal single-mode contribution $\ln2$.

The entropy measures the extent to which the occupied Bloch subspace fails to
factorize across the unit-cell partition.
If a single compact orbital spans the target manifold,
$\lambda_1=1$ and $\lambda_{a>1}=0$, implying
$S_{\rm cell}=0$.
For Chern bands this entanglement cannot be removed, implying
$S_{\rm cell}>0$ for every choice of partition
(Sec.~\ref{subsec:irremovable_zeros_topological}).

The converse does not hold. A topologically trivial band also generically gives
$S_{\rm cell}>0$.
Since Eq.~\eqref{eq:sumrule} fixes the total occupation
independently of where the cell boundary is drawn, moving the partition
center can only redistribute that weight among the CNOs, and the entropy
falls as the weight concentrates into fewer modes. Minimizing
$S_{\rm cell}$ over partitions is therefore equivalent to
selecting the center(s) at which the smallest number of compact orbitals
overlaps with target bands.

\section{Consequences of Symmetry and Topology}\label{subsec:constraints_CNO_symmetry_topology}

The previous section established that the CNOs and their occupations depend on the real-space partition. We now show that this choice also fixes the symmetry label of each CNO, and that this label controls where the projected envelope function $s_{a,\bk}$ has zeros in the Brillouin zone.

\subsection{Symmetries and CNOs}
\label{subsec:symmetry_reps_of_CNOs}

A set of bands describes an atomic insulator only if its band representation decomposes into a sum of EBRs~\cite{Zak1982,Cano2018}. If no such decomposition exists, the bands cannot be represented by any symmetric collection of localized orbitals and therefore possess a topological obstruction \cite{Xu2020OAI, Wieder2022TopologicalSymmetry}.
Determining whether such a decomposition exists is generally nontrivial, but systematic methods are now available~\cite{PoVishwanathWatanabe2017}.

CNOs transform as irreducible representations (irreps) of the site-symmetry group $G_q$ respected by the Wyckoff position $q$ around which the chosen unit cell partition is centered\footnote{The symmetry group respected by the partition is the site symmetry group of its center, \textit{not} modulo translations. This group is spanned by the point group symmetries that map every position contained within the chosen unit cell onto another position of the same unit cell, and not a neighboring cell. This does not affect continuum models, but, in tight-binding models with orbitals at the boundaries of the unit cell, it implies the breaking of the symmetries that map these orbitals to equivalent positions but in neighboring unit cells. This is analogous to the symmetry properties of cluster DMFT.}. We denote by $L_{\bR,q}$ the UC-1pRDM centered at this Wyckoff position $q$ in unit cell $\bR$, whose eigenstates satisfy
\begin{equation}
    L_{\bR,q}|\tau_{\bR,a_q}\rangle
    =
    \lambda_{a_q}
|\tau_{\bR,a_q}\rangle.
\end{equation}
Each CNO carries the composite label $a_q$, where $a$ labels the mode associated with the partition centered at $q$, which transforms according to the irrep $\rho_a$ of $G_q$. Translating $|\tau_{\bR, a_q}\rangle$ to every unit cell gives the CNO Bloch state
\begin{equation}
    |\tau_{\mathbf k,a_q}\rangle = \dfrac{1}{\sqrt{N_k}} \sum_{\mathbf R} e^{i\mathbf k\cdot(\mathbf R+\mathbf r_q)} |\tau_{\mathbf R,a_q}\rangle,
\label{eq:CNO_bloch_sum}
\end{equation}
which realizes the EBR induced from $(q,\rho)$.
The CNOs $|\tau_{a_q}\rangle$ therefore provide an optimal choice of orbitals for the $(q,\rho)$ EBR.

\subsection{Symmetry-Enforced Zeros}
\label{subsec:removable_zeros_OAI}

Symmetry alone can force the overlap between CNO and target Bloch states to vanish.
At momentum $\mathbf k$, both the target Bloch states and the CNO Bloch states furnish representations of the little group $G_{\mathbf k}$.
If they transform according to different irreps of $G_{\mathbf k}$, Schur's lemma implies that $\langle\psi_{\mathbf k n}|\tau_{\mathbf k,a_q}\rangle=0$.
Consequently, any candidate EBR decomposition must first reproduce the little-group irreps of the target bands before it can possibly span the desired subspace.
Matching symmetry representations is therefore a necessary, but not sufficient, condition for constructing a local model.

\begin{figure}
    \centering
    \includegraphics[width=0.9\linewidth]{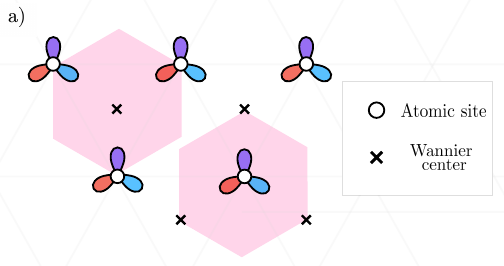}
    \includegraphics[width=0.9\linewidth]{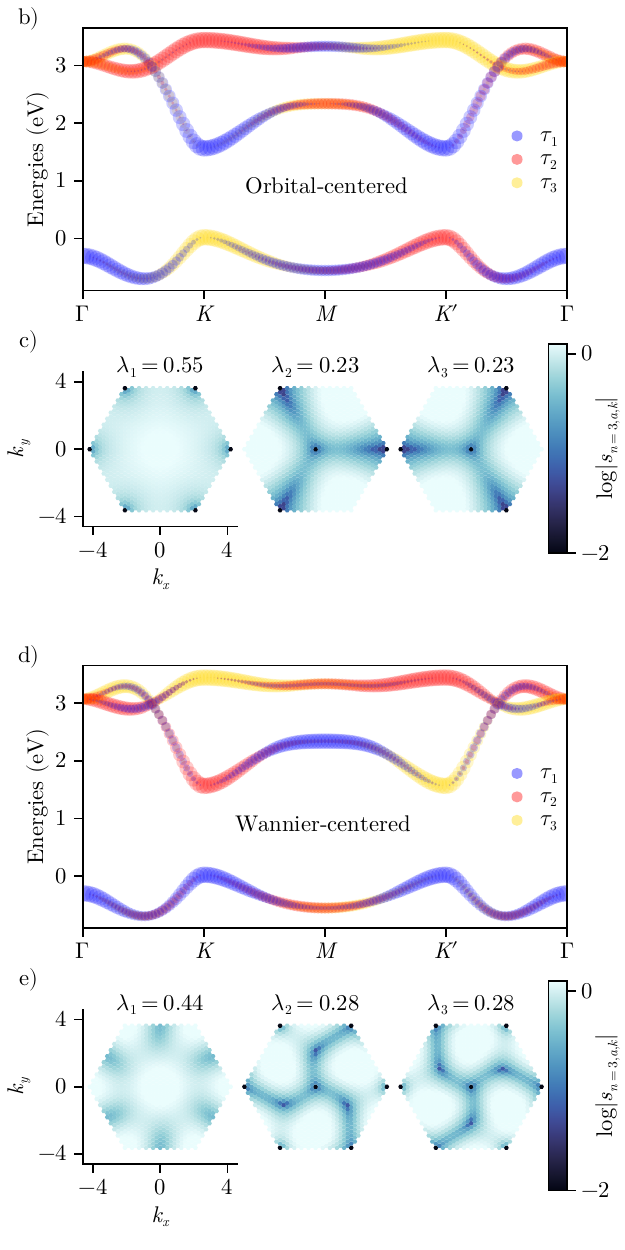}
    \caption{
    a) Triangular-lattice three-orbital monolayer TMD model. The $1a$ Wyckoff position where the $d$ orbitals are centered is indicated by the white circles. These $d$ orbitals can be combined into a basis of three $sp_2$-like orbitals. Black crosses mark the $1c$ Wannier centers, where the $sp_2$ orbitals from three neighboring cells constructively interfere. The two shaded cells show the partition choices. 
    b-d) CNOs and envelope function of the valence band of monolayer WSe$_2$ for a-b) $1a$-centered unit cell, c-d) $1c$-centered unit cell. In a) and c), the band structure of the 3-band model is colored by the overlaps $| \langle u_{\bk n} | \tau_a \rangle|$ for each CNO. b) and d) show the envelope functions $s_{a, \bk}$ over the Brillouin zone}
    \label{fig:TMD_CNOs}
\end{figure}

Obstructed atomic insulators are insulators which admit an exponentially localized Wannier function, but its center does not coincide with any atomic site \cite{Cano2022_oai}.
The band representation induced from the atomic orbitals does not match the band representation of the occupied bands, signaling that the electrons have reorganized into a bonding pattern incompatible with the underlying atomic positions.

We illustrate this through the three-orbital model of monolayer $1H$-TMDs of Ref.~\cite{DiXiao2013}, which are obstructed atomic insulators protected by $C_{3z}$-symmetry~\cite{Zeng2021MultiorbitalDichalcogenides,Qian2022CnDichalcogenides,Jung2022HiddenDichalcogenides,Eck2023RecipeLattice,HolbrookIngham2026}. Their valence band is formed primarily from the metal $\{|d_{z^2} \rangle,  |d_{x,y} \rangle, |d_{x^2 - y^2} \rangle\}$ orbitals, which form a triangular lattice ($1a$ Wyckoff position, open circles in Fig.~\ref{fig:TMD_CNOs}(a)). Transforming these three $d$ orbitals into three $sp_2$-like orbitals shows that the main hopping process occurs between the three $sp_2$-like orbitals pointing towards the center of the down-pointing triangles, misaligned with both the metal and chalcogen sites (1c Wyckoff position, crosses in Fig.~\ref{fig:TMD_CNOs}a)). This constructive interference leads to a different character under  $C_{3z}$ at the $\Gamma$ and $K/K'$ points, and therefore to the obstruction: the band has $|d_{z^2} \rangle$ character at $\Gamma$ and is therefore trivial under $C_{3z}$, whereas it has $|d_{x,y} \rangle \mp i |d_{x^2 - y^2} \rangle$ character at $K/K'$ and thus gets a phase $e^{i\pm2\pi/3}$ under $C_{3z}$. Therefore, this band can only be Wannierized by a $s$-orbital (trivial under $C_{3z}$) centered at the 1c Wyckoff position.

Here, we study the CNO spectrum of the valence band, and study how the choice of unit cell partition reflects the obstruction. In particular, we consider the two unit cell partitions shown as pink hexagons in Fig.~\ref{fig:TMD_CNOs}a), centered at the $1a$ (circles) and $1c$ (crosses) Wyckoff positions, respectively. 

The $1a$-centered unit cell spans the three $d$/$sp_2$ orbitals centered at the same $1a$ position. Due to the $C_{3z}$ symmetry, the CNOs are exactly the original $d$ orbitals, $|\tau_1 \rangle = |d_{z^2} \rangle$, $|\tau_{2,3} \rangle = (|d_{x,y} \rangle \mp i |d_{x^2 - y^2} \rangle)/\sqrt{2}$, with $|\tau_1 \rangle = |d_{z^2} \rangle$ dominating the spectral weight ($\lambda_1 = 0.55, \lambda_{2,3} = 0.23$). The obstruction forces zeros in the envelope function $s_{a, \bk} = \langle u_{\bk, \textrm{val}} | \tau_a \rangle/\sqrt{\lambda_a}$ for all the CNOs at two of the $\Gamma$, $K$ and $K'$ points (see Fig.~\ref{fig:TMD_CNOs}b),c)).

The $1c$-centered unit cell considers the three $sp_2$ orbitals that point to the same $1c$ position but are located in three different $1a$ positions. The leading $1c$-centered CNO is the $C_3$ trivial combination and has no zeros in its envelope function (see Fig.~\ref{fig:TMD_CNOs}d),e)), indicating that the valence band of monolayer WSe$_2$ can be fully spanned by a single local orbital at the $1c$ Wyckoff position.

\subsection{Topology-Enforced Zeros}
\label{subsec:irremovable_zeros_topological}
For a band with finite Chern number, say $C = 1$, there cannot exist a smooth, nowhere-vanishing section of the Bloch bundle over the torus \cite{BrouderPRL2007, Monaco2018}. 
Since $|\tau_1\rangle$ is a fixed vector in orbital space, the overlap function $\langle u^\mathrm{per}_{\bk}|\tau_1\rangle$ is a smooth map from the torus to $\mathbb{C}$.
The topological obstruction guarantees the existence of at least one momentum $\bk_0$ where $\langle u^\mathrm{per}_{\bk_0}|\tau_1\rangle 
= 0$ \cite{Mera2022, ThonhauserVanderbilt2006}.

\begin{figure}
    \centering
    \includegraphics[width=\linewidth]{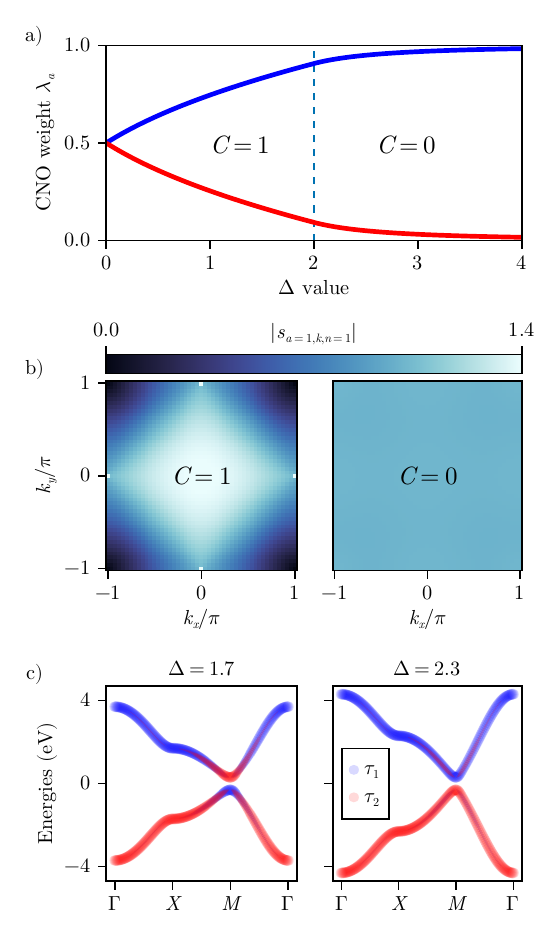}
    \caption{a) CNO weights $\lambda_a$ for the top band of the spinless BHZ model. The topological transition occurs at $\Delta = 2$, shown by the vertical dashed line. b) Absolute value of the envelope function of the dominant CNO $|s_{a=1}|$ along the Brillouin zone for the left: $C=1$ regime and right: $C=0$ regime. In the $C=1$ regime, $s_{a=1}$ goes to zero at the $M$ points, while in the $C=0$ case, the envelope function is finite everywhere. c) band structure along a high-symmetry path in left: the $C=1$ regime and right: the $C=0$ regime. The points are colored by CNO weights obtained from the UC-1pRDM of the top band.}
    \label{fig:BHZ_model}
\end{figure}

We illustrate these ideas with the spinless Bernevig-Hughes-Zhang (BHZ) model, which is the simplest lattice realization of a Chern insulator with a tunable phase transition. The model has two orbitals per unit cell with $s$ and $p_z$ character, both centered at the origin of the unit cell. The $\bk$-space Hamiltonian is $H_\bk = \mathbf{h}_\bk\cdot
\boldsymbol{\sigma}$, with
\begin{equation}
\mathbf{h}_\bk
=
\begin{pmatrix}
\sin k_x \\
\sin k_y \\
\Delta + \cos k_x + \cos k_y
\end{pmatrix}
\label{eq:BHZ}
\end{equation}
and $\boldsymbol{\sigma} = (\sigma_x, \sigma_y, \sigma_z)$.
The model has a single parameter $\Delta$ that drives a topological phase transition. The top band has Chern number $|C| = 1$ for $|\Delta| < 2$ and $C = 0$ for $|\Delta| > 2$, with gapless points at $\Delta = \pm 2$. For each value of $\Delta$, we compute the CNO spectrum for the unit cell centered at the origin.

Fig.~\ref{fig:BHZ_model}a) shows the two CNO occupations as a function of $\Delta$. 
In the trivial phase ($|\Delta| > 2$), $\lambda_1$ (blue line) approaches unity and $\lambda_2$ (red line) approaches zero, reflecting the fact that the top band becomes increasingly $s$-like or $p_z$-like as $\Delta \to \pm\infty$. 
A single CNO is sufficient to represent the band, consistent with the existence of exponentially localized Wannier functions. 
At the phase boundaries $\Delta = \pm 2$, the gap closes and the occupation spectrum changes slope but remains continuous. 
In the topological phase ($|\Delta| < 2$), both occupations are fractional. 
At $\Delta = 0$, the two occupations equalize at $\lambda_1 = \lambda_2 = 1/2$, giving maximal unit-cell entanglement $S_\mathrm{cell} = \ln 2$ per mode.

Fig.~\ref{fig:BHZ_model}a) makes concrete the point raised at the start of this section. The occupations vary smoothly through the transition and do not by themselves separate the two phases, while the envelope function $s_{1,\bk}$ does.
Fig.~\ref{fig:BHZ_model}b) shows the absolute value of the envelope function $|s_{1,\bk}|$ over the Brillouin zone for representative values of $\Delta$ in the two phases.
In the trivial phase, $|s_{1,\bk}|$ is smooth and nonvanishing everywhere.
In the topological phase, $|s_{1,\bk}|$ vanishes at the $M$ point, but, as established above, its existence is protected. The band therefore requires both CNOs, and no single local orbital can span it.

\subsection{Revealing Band Inversions}\label{subsec:band_inversion}
While the CNOs are defined from the bands of interest (in this case, the top band), their projections onto other
Bloch bands can reveal nearby band inversions. 
At fixed $\bk$, completeness of the Bloch eigenvectors gives
\begin{equation}
    \sum_{n} | \langle u^{\textrm{per}}_{\mathbf{k} n}| \tau_a \rangle |^2 = 1 
\end{equation}
for any normalized state $|\tau_a \rangle$.
For a two-band model such as the BHZ model, it is straightforward to deduce that for any point in the BZ, any remaining weight of a CNO must live on the second band.
This is illustrated in Fig.~\ref{fig:BHZ_model}c) for two representative values of $\Delta$. 
On the right panel, the band structure of the BHZ model in the trivial phase is colored by the CNO weights $|\langle u^{\textrm{per}}_{\bk n}| \tau_a \rangle|$ for both bands and both CNOs.
We see that $\tau_1$ has finite weight across the entirety of the top band, with some variance close to the $M$ point where it lives partially in the second band.
Once in the topological phase however, the first CNO has zero weight at the $M$ point of the top band as expected and instead lives solely in the bottom band at the $M$ point.

Importantly, this orbital decomposition applies to continuum models. Consequently, the CNO projections can identify band inversions and guide the construction of effective real-space models.

\section{Constructing Local Models from CNOs}\label{sec:constructing_TB}
This section outlines the construction of local models from CNOs.
Each CNO is a compact orbital associated with a Wyckoff position.
The combined overlap matrix determines whether this set spans the target bands, while the total spectral weight ranks competing choices with the same number of orbitals.
\subsection{Optimal Overlap}\label{sec:overlap}

In general, multiple EBR sets can reproduce the irreps of the target bands at high-symmetry points, and these candidate sets may contain orbitals centered on different Wyckoff positions.
Beyond symmetry, the chosen EBR set must completely span the target subspace.

For a given set of EBRs $\mathcal{A}$, the overlap of the CNO Bloch states with the target bands defines the Gram matrix
\begin{align}
    \left[ \mathcal{S}_{\bk}\right]_{n,n'}
    &=
    \sum_{a_q \in \mathcal{A}}
    \langle \psi_{\bk n} | \tau_{\bk,a_q} \rangle
    \langle \tau_{\bk,a_q} | \psi_{\bk n'} \rangle
    \nonumber \\
    &=
    \sum_{a_q \in \mathcal{A}}
    \lambda_{a_q}
    s_{a_q,n,\bk}
    s^*_{a_q,n',\bk},
    \label{eq:gram_matrix_S}
\end{align}
where $n,n'\in P$.
A zero in $\det \mathcal{S}_{\bk}$ at any momentum in the Brillouin zone indicates that the EBR set fails to span the target bands.

Among all EBR sets that satisfy this completeness condition, we rank candidate local models using both the total spectral weight of the CNO trials and the resulting Wannier spread.
For a single choice of partition, the spectral weight of an individual CNO is given by its occupation $\lambda_{a_q}$.
When CNOs from different partitions are combined, the total spectral weight is instead defined as
\begin{equation}
    \lambda_{\mathcal{A}}
    = \frac{1}{N_k}
    \Tr(PP_{\mathcal{A}}),
    \label{eq:spectral_weight_general}
\end{equation}
where $P$ is the projector onto the target bands defined in Eq.~\eqref{eq:def:P_M}, and the projector onto the (nonorthogonal) CNO subspace is
\begin{equation}
    P_{\mathcal{A}}
    =
    \sum_{\bk}
    \sum_{a_q,a_q'\in\mathcal{A}}
    |\tau_{\bk,a_q}\rangle
    \left[\mathcal{O}_{\bk}^{-1}\right]_{a_q,a_q'}
    \langle\tau_{\bk,a_q'}|.
\end{equation}
Equation~\eqref{eq:spectral_weight_general} is normalized such that it is consistent with Eq.~\eqref{eq:sumrule}.
Since CNOs obtained from different partitions are not necessarily orthogonal, the overlap matrix
\begin{equation}
    \left[\mathcal{O}_{\bk}\right]_{a_q,a_q'}
    =
    \langle\tau_{\bk,a_q}|\tau_{\bk,a_q'}\rangle
\end{equation}
is required to construct the projector onto the CNO Bloch states.
For an EBR set composed of CNOs from a single partition $q$, Eq.~\eqref{eq:spectral_weight_general} reduces to the sum of the individual occupations,
$\lambda_{\mathcal{A}}=\sum_{a_q}\lambda_{a_q}$.

\subsection{Single-Shot Wannierization}
\label{subsec:wannier}

Our discussion so far has focused on selecting a set of CNOs; turning that set into a tight-binding model requires one further step. When the selected CNOs span the target bands at every $\bk$, they can serve as trial states for Wannierization \cite{MarzariMLWFRMP2012}, which retains the target bands exactly and uses the CNOs only to fix the gauge.
We briefly review the single-shot Wannierization procedure based on projections from initial trial states, following Ref.~\cite{SouzaMarzariVanderbilt2001}.

A Wannier function in $\bk$ space is obtained from a unitary transformation of the target Bloch states,
\begin{equation}
|w_{\bk a}\rangle
=
\sum_{n\in P}
M_{\bk,na}
|\psi_{\bk n}\rangle,
\label{eq:wannier_def}
\end{equation}
where the mixing matrix $M_\bk$ specifies the gauge choice.
For a target projector $P$ with $N_P$ bands and a set of $N_\tau$ trial states $\{|g_a\rangle\}$, with generically $N_\tau\geq N_P$, the overlap between Bloch states and trial states is collected into the matrix
\begin{equation}
    A_{\bk,ma}
    =
    \eta_{\bk m}
    \langle\psi_{\bk m}|g_a\rangle.
\end{equation}
The factor $\eta_{\bk m}$ is the weighting factor coming from the projection onto the desired energy window that allows a smooth disentanglement~\cite{SouzaMarzariVanderbilt2001}.
A singular value decomposition of this matrix,
\begin{equation}
    A_\bk
    =
    U_\bk\Sigma_\bk W_\bk^\dagger,
\end{equation}
defines the mixing matrix appearing in Eq.~\eqref{eq:wannier_def} as
\begin{equation}
    M_\bk
    =
    U_\bk W_\bk^\dagger.
\end{equation}
For the Wannier functions to be exponentially localized, the transformation between Bloch states and Wannier functions must remain unitary throughout the Brillouin zone.
This requires that $A_\bk$ is square and nonsingular at every $\bk$.

When $N_\tau=N_P$, the disentanglement weighting is chosen as $\eta_{\bk m}=1$ for $m\in P$ and $\eta_{\bk m}=0$ otherwise.
Together with the condition
$\det\mathcal{S}_\bk\neq0$
at all $\bk$, this ensures that $A_\bk$ is square and invertible.
This scheme should result in exponentially localized Wannier functions as long as the projector does not have a Chern number.

In the case of topological bands, we are forced to work with $N_\tau>N_P$. The resulting trial-state space is larger than the target subspace and the disentanglement weighting $\eta_{\bk m}$ can include Bloch states outside of $P$.
In this case, we can consider $\eta_{\bk m}=1$ for $m\in P$ and guess $\eta_{\bk m}$ for $m\notin P$ as a suitable normalized energy-dependent decaying function~\cite{Crepel2024_prr,Fischer2025TheoryTWSe2,Munoz-Segovia2025Twist-angleWSe2}.
The precise form of this function does not control whether a local effective model can be constructed, but it provides a tunable parameter for optimizing the localization of Wannier functions.
With the band inversion identification discussed in Sec.~\ref{subsec:band_inversion}, CNOs help us make an educated guess for the function $\eta_{\bk m}$.

Once the Wannier functions are fixed, the tight-binding parameters follow directly
\begin{equation}
    t_{ab}(\bR) = \langle w_{\bR a} | H | w_{\mathbf{0} b} \rangle \, ,
    \label{eq:hoppings}
\end{equation}
from which one can construct the effective tight-binding model.

\subsection{Real-Space Grid from Plane-Wave Basis}
\label{subsec:real_space_grid_plane_wave}

Continuum models, which are the natural starting point for {\moire} systems are represented in a plane-wave basis and do not contain an \textit{a priori} notion of orbital positions within the unit cell.
Consider a continuum Hamiltonian written in the plane-wave basis,
\begin{equation}
H
=
\sum_\bk
\sum_{\substack{\bG,\bG'\\ l,l'}}
c^\dagger_{\bk+\bG,l}
[H(\bk)]_{(\bG,l),(\bG',l')}
c^{\vphantom{\dagger}}_{\bk+\bG',l'},
\label{eq:continuum_H}
\end{equation}
where $\bG$ denotes moiré reciprocal lattice vectors and $l$ collects internal degrees of freedom such as layer, valley, and spin.
Diagonalizing $H(\bk)$ gives $H(\bk)|\psi_{\bk n}\rangle
=
\varepsilon_{\bk n}|\psi_{\bk n}\rangle$
 with
\begin{equation}
|\psi_{\bk n}\rangle
=
\sum_{\bG,l}
z^n_{\bk,\bG,l}
|\bk+\bG,l\rangle .
\label{eq:continuum_eigenvectors}
\end{equation}
\begin{figure}
    \centering
    \includegraphics[width=\linewidth]{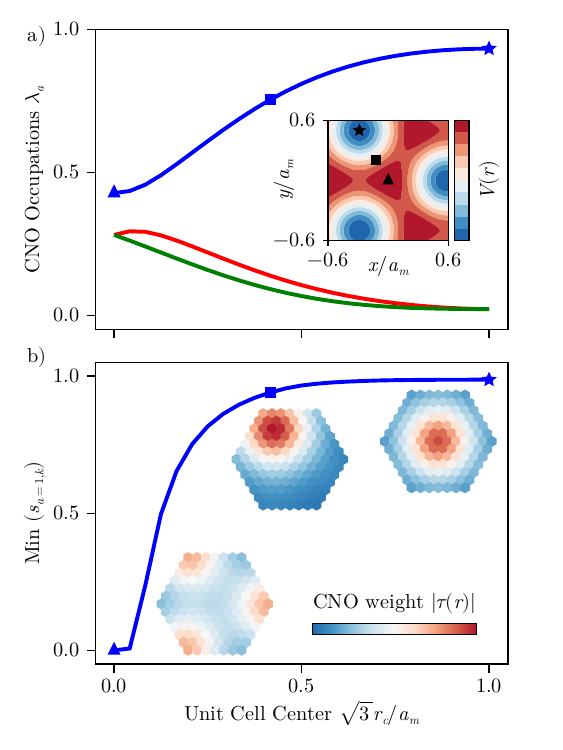}
    \caption{a) CNO occupation spectra for the lowest band of a single parabolic band in a triangular {\moire} potential as a function of scaled unit-cell center $r_c$. The inset shows the real-space potential $V(\br)$ and three highlighted partition centers. b) Minimum value of the dominant envelope function $|s_{1,\bk}|$ as a function of $r_c$, with representative real-space profiles of the leading CNO.}
    \label{fig:triangular_moire_sliding}
\end{figure}

To define the cell-periodic states and the embedding matrix, we Fourier transform the reciprocal basis vectors $|\bG\rangle$ to their real-space dual grid inside the unit cell:
\begin{equation}
|\phi_{\br_\alpha l}\rangle
=
\frac{1}{\sqrt{N_k}}
\sum_{\bR}
|\bR+\br_\alpha,l\rangle ,
\label{eq:continuum_delta_basis}
\end{equation}
where $\bR$ is a moiré lattice vector and $\br_\alpha$ labels 
points within a single unit cell.
The UC-1pRDM in Eq.~\eqref{eq:ucRDM_matrix} is then defined using
\begin{equation}
V_\bk
|u_{\bk n}\rangle
=
\sum_{\br_\alpha,l}
e^{i\bk\cdot\br_\alpha}
z^n_{\bk,\br_\alpha,l}
|\phi_{\br_\alpha l}\rangle ,
\label{eq:continuum_per}
\end{equation}
with the real-space coefficients given by
\begin{equation}
z^n_{\bk,\br_\alpha,l}
=
\frac{1}{\sqrt{N_G}}
\sum_{\bG}
e^{i\bG\cdot\br_\alpha}
z^n_{\bk,\bG,l}.
\label{eq:FT_eigvec}
\end{equation}

Formally, $\br_\alpha$ is a continuous coordinate inside the unit cell, but a finite reciprocal cutoff provides only $N_G$ independent real-space degrees of freedom for each internal index.
In this basis, the continuum unit-cell one-particle reduced density matrix remains defined by Eq.~\eqref{eq:ucRDM_matrix}.
It acts on the $N_GN_l$-dimensional Hilbert space spanned by $|\phi_{\br_\alpha l}\rangle$, and its eigenvectors are compact CNOs defined on the real-space grid.
The cutoff $N_G$ controls the spatial resolution of the resulting local orbitals and must be converged.

\subsection{Parabolic Band in a Hexagonal {\Moire} Potential}
\label{subsec:triangular_moire}

We consider a parabolic band in a hexagonal {\moire} potential with the Hamiltonian
\begin{equation}
H
=
\frac{\hbar^2\bk^2}{2m^*}
+2V\sum_{j=1}^3
\cos(\bG_j\cdot\br+\phi),
\label{eq:triangular_moire_H}
\end{equation}
where the three $\bG_j$ are related by $C_3$. We use $V=50\,\textrm{meV}$ and $\phi=60^\circ$, giving a triangular potential with a minimum at the position marked by a star in the inset of Fig.~\ref{fig:triangular_moire_sliding}a). The lowest band has Chern number $C=0$, and its projected charge density localizes near the potential minima, away from the origin.

The real-space partition is the hexagonal Wigner-Seitz cell of the moiré lattice, but its center $r_c$ can be shifted. 
Only the partitions centered on high-symmetry extrema of the potential preserve $C_3$, which are the maximal Wyckoff positions at $1a$ and $2b$. Fig.~\ref{fig:triangular_moire_sliding}a) shows the CNO occupations as $r_c$ is moved between representative high-symmetry centers. 
Across the sweep there is a dominant CNO, while the subleading modes become degenerate only when the partition preserves $C_3$.

Figure~\ref{fig:triangular_moire_sliding}b) tracks the minimum of the envelope function $|s_{1,\bk}|$. 
When the partition is centered on a potential maximum, and hence away from the charge center of the lowest band, the leading occupation is sizable but the envelope has a symmetry-enforced zero at the $K$ and $K'$ points.
This is the continuum analogue of the obstructed atomic examples above where the unit cell has split the charge center. 
When the partition is centered on the potential minimum, $\lambda_1$ increases to roughly $0.93$ and the minimum of $|s_{1,\bk}|$ is close to unity, showing that a single compact CNO captures the lowest band to high accuracy.

The CNO center is therefore a physical variational choice. When several symmetry-preserving Wyckoff positions are available, moving the partition can reduce the number of local orbitals needed.
Equivalently, the sweep in Fig.~\ref{fig:triangular_moire_sliding}a) minimizes the unit-cell entanglement entropy of Eq.~\eqref{eq:Scell} over the partition center.

\section{Twisted Transition Metal Dichalcogenides}
\label{sec:tWSe2}

In recent years, twisted transition metal dichalcogenides have emerged as a tunable platform to explore a variety of correlated and topological phases such as Mott and magnetic insulators \cite{wang_correlated_2020,Tang2020SimulationSuperlattices,Anderson2023,ghiotto_stoner_2024,knuppel_mak_correlated_2024}, generalized Wigner crystals \cite{xu_correlated_2020,Regan2020MottSuperlattices,huang_correlated_2021,Li2021ImagingCrystals}, topological insulators \cite{Li2021QuantumBands,tao2024valley,Foutty2024MappingWSe2}, Kondo insulators and heavy Fermi liquids \cite{zhao_gate-tunable_2023,zhao_emergence_2024,han_topological_2026,tian_emergence_2026}, and superconducting phases \cite{Xia2024SuperconductivityWSe2,Guo2025SuperconductivityWSe2,Xia2026Bandwidth-tunedWSe2,Guo2026AngleWSe2}. 
The theoretical exploration of these systems is often facilitated by building lattice models with manageably few degrees of freedom \cite{Crepel2024_prr, Qiu_Fengcheng_PRX}.
Methods such as Dynamical Mean Field Theory (DMFT)~\cite{VollhardtMetzner1989, RozenbergGeorges1996} and functional Renormalization Group (fRG)~\cite{Salmhofer2001FermionicTheory,Metzner2012FunctionalSystems,Platt2013FunctionalInstabilities,Dupuis2021TheApplications,Beyer2022ReferenceGroup}, for example, rely on accurate real-space models in order to treat the many-body problem.
When the starting point is a continuum model, Wannierization is a common technique to generate an atomistic model and here, the CNO construction is helpful in two ways.
The first is accurately assessing the quantum geometry of the bands even before any attempt at a local model, and the second is providing trial states for an effective local model even for a topologically nontrivial set of bands.

We construct the CNOs for the continuum model of twisted bilayer WSe$_2$~\cite{Wu_firstTMD} using the parameters of Ref.~\cite{DevakulCrepelFu2021},
$(V,w,\psi,m^*)=(9.0\,\textrm{meV}, 18.0\,\textrm{meV}, 128^\circ, 0.43m_e)$ and focus on a twist angle range of $\theta\in[1^\circ,5^\circ]$.
Each spin-valley sector is described by a single parabolic band in a $C_3$-symmetric moiré potential. 
All results below are quoted for one valley; the opposite valley follows by time reversal, with the Chern numbers reversed in sign.
For this parameter set, the top two {\moire} bands can be divided into two regimes. 
For the range $1.51^\circ < \theta \leq 5.0^\circ$, the top two moiré bands both carry Chern number $C=-1$.
At $\theta \sim 1.51^\circ$, the second and third bands touch at the $\Gamma$ point and a band inversion occurs, leading to the second band switching Chern number to $C = 1$ for $\theta < 1.51^\circ$, and the top two bands having a total Chern number of zero. 

\subsection{Twist Angle Evolution of CNO spectrum}
\begin{figure}
    \centering
    \includegraphics[width=\linewidth]{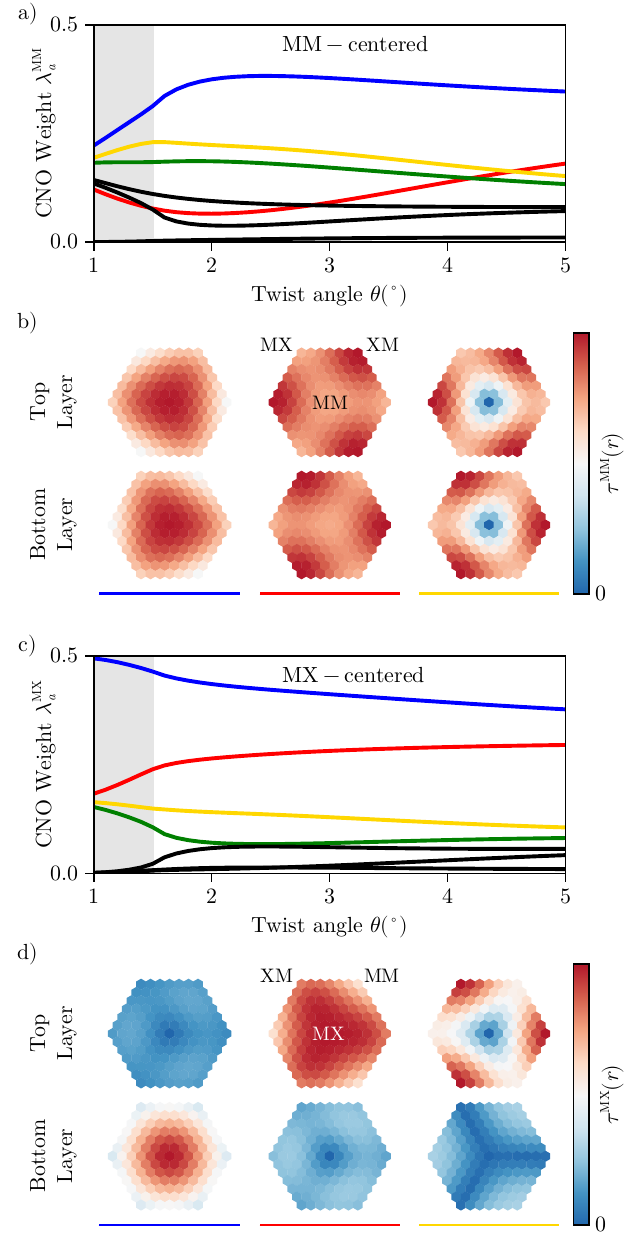}
    \caption{a), c) CNO occupations $\lambda^q_a$ for the top two bands of continuum tWSe$_2$ as a function of twist angle $\theta$ for partition centers $q =$ MM and $q=$ MX, respectively. 
    b), d) Representative real-space CNO profiles in the topological regime for the two partitions. The leading CNO for the MM-centered partition localizes near the MM stacking region, while subleading modes carry weight on the surrounding MX and XM regions.}
    \label{fig:CNOs_continuum_twse2_real_space}
\end{figure}
We compute the UC-1pRDM of the top two bands in two Wigner-Seitz unit cells centered at MM and MX sites. 
Both the MM- and MX-centered partitions preserve $C_{3z}$, and only the MM-centered partition preserves $C_{2y}$ symmetry, which relates the two layers.
$C_{2y}$ symmetry relates the MX-centered CNOs to the XM-centered CNOs so in order to preserve $C_{2y}$, any MX-centered CNO must be considered in tandem with its XM-centered pair.

%

Figure~\ref{fig:CNOs_continuum_twse2_real_space}a) and c) shows the CNO weights for the MM- and MX-centered partitions respectively, across the twist angle range $\theta\in[1^\circ,5^\circ]$. 
The line colors label each CNO across angles. 
The grey shadowed region in Fig.~\ref{fig:CNOs_continuum_twse2_real_space}a) and c) indicates the topologically trivial regime in which the total Chern number of the top two bands is zero.
The real-space profile of representative CNOs from the topologically nontrivial regime $\theta > 1.51^\circ$ are plotted in Fig.~\ref{fig:CNOs_continuum_twse2_real_space}b) and d) for the two partition centers respectively.

In the topologically nontrivial regime, the finite total Chern number of the top two bands precludes any local state from completely spanning the band.
Despite this, there is a hierarchy in CNO weights since $|\tau^{\textrm{MM}}_1 \rangle$ is a clearly dominant mode across most of the range, indicating that it has the most overlap with the subset of interest.
As seen in Fig.~\ref{fig:CNOs_continuum_twse2_real_space}b), this leading CNO is localized on the MM stacking region.
The subleading modes are generically localized away from the MM site, suggesting that the MX/XM-centered partitions should be considered.
Indeed, the UC-1pRDM with the MX-centered partition yields a leading CNO with a dominant weight localized on the MX site in one layer (Fig.~\ref{fig:CNOs_continuum_twse2_real_space}d)).
The overlap between the topmost band and the leading CNO in the MM-centered partition generically has a zero pinned to $\Gamma$, while the one with the leading CNO in the XM(MX)-centered has zeros at the $K (K')$ points. These pinned locations follow from the little-group characters (see Table~\ref{tab:CNO_band_characters}).
\begin{figure}
    \centering
    \includegraphics[width=\linewidth]{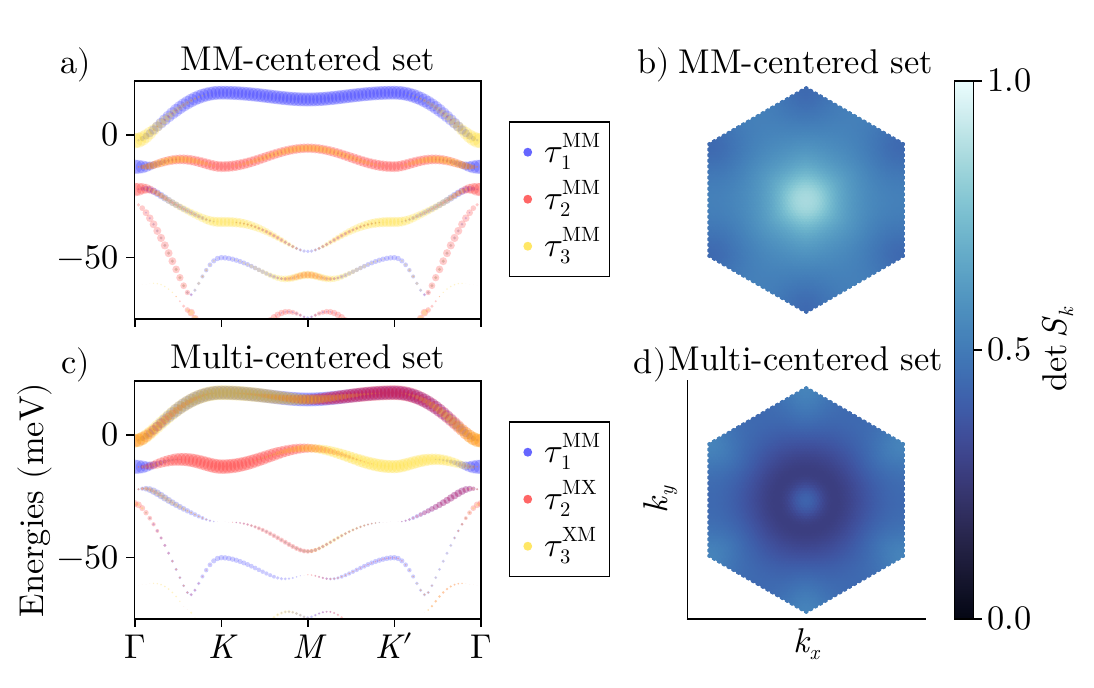}
    \caption{a) Band structure of $3.0^\circ$ twisted WSe$_2$ colored by the weights of the top three CNOs computed from the top two bands, using an MM-centered unit cell. b) Determinant of the combined overlap matrix $\det \mathcal{S}_\bk$ over the Brillouin zone for the MM-centered CNO set $\{ | \tau^{\textrm{MM}}_1 \rangle, | \tau^{\textrm{MM}}_2 \rangle, | \tau^{\textrm{MM}}_3 \rangle \}$. c) Band structure of $3.0^\circ$ twisted WSe$_2$ colored by the weights of the dominant CNO computed from the top two bands, using an MM-, MX-, and XM-centered unit cell. d) Determinant of the combined overlap matrix $\det \mathcal{S}_\bk$ over the Brillouin zone for the multi-centered CNO set $\{ | \tau^{\textrm{XM}}_1 \rangle, | \tau^{\textrm{MM}}_1 \rangle, | \tau^{\textrm{MX}}_1 \rangle \}$.}
\label{fig:twse2_top_two_bands_det_Sk}
\end{figure}

At $\theta \sim 1.51^\circ$, the second and third bands touch and a band inversion at $\Gamma$ occurs where the top two bands switch their total Chern number to zero. 
As in Sec.~\ref{subsec:irremovable_zeros_topological}, the CNO weights themselves vary smoothly across the topological phase transition, as seen in the grey regions of Fig.~\ref{fig:CNOs_continuum_twse2_real_space}a) and c). 
The change in topology is reflected instead in the envelope function of the second band, where the zero is lifted and a local state is able to fully span both bands. 
What the CNO construction tells us about this regime can be seen in the dominant CNO weight in Fig.~\ref{fig:CNOs_continuum_twse2_real_space}c) as $\lambda^{\textrm{MX}}_1 \rightarrow 0.5$. The top two bands form a hexagonal lattice and can be described almost exactly with two orbitals, one centered at the MX stacking site, and the other at the XM site.

\subsection{Local Real-Space Model}\label{subsec:tWSe2_TBmodel}

As an illustrative example of the construction of a local model from CNOs, we consider twisted WSe$_2$ at a twist angle of $\theta = 3^\circ$.
The band structure is shown in Fig.~\ref{fig:twse2_top_two_bands_det_Sk} with the CNO projections $\langle u^\textrm{per}_{\bk n} | \tau_a \rangle$ for the dominant three CNOs in the MM-centered partition (Fig.~\ref{fig:twse2_top_two_bands_det_Sk}a)) and in the MX-centered partition (Fig.~\ref{fig:twse2_top_two_bands_det_Sk}c)).

\begin{figure*}
    \centering
    \includegraphics[width=2\columnwidth]{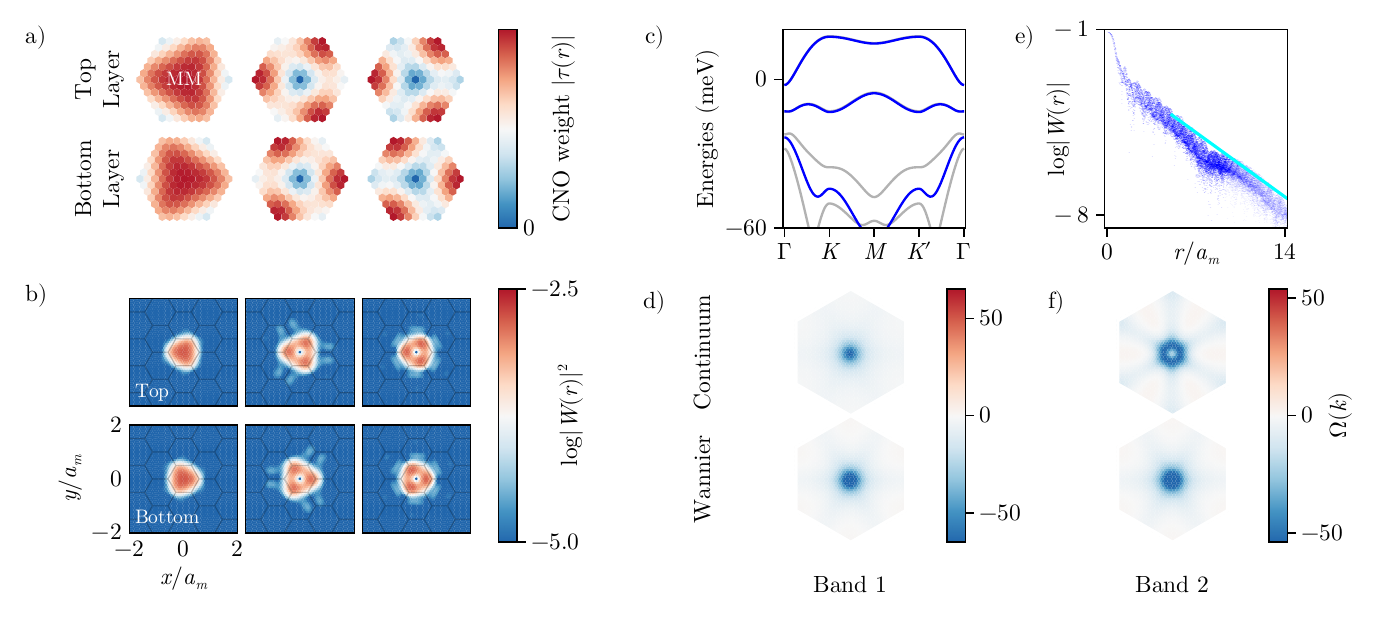}
    \hrule
    \includegraphics[width=2\columnwidth]{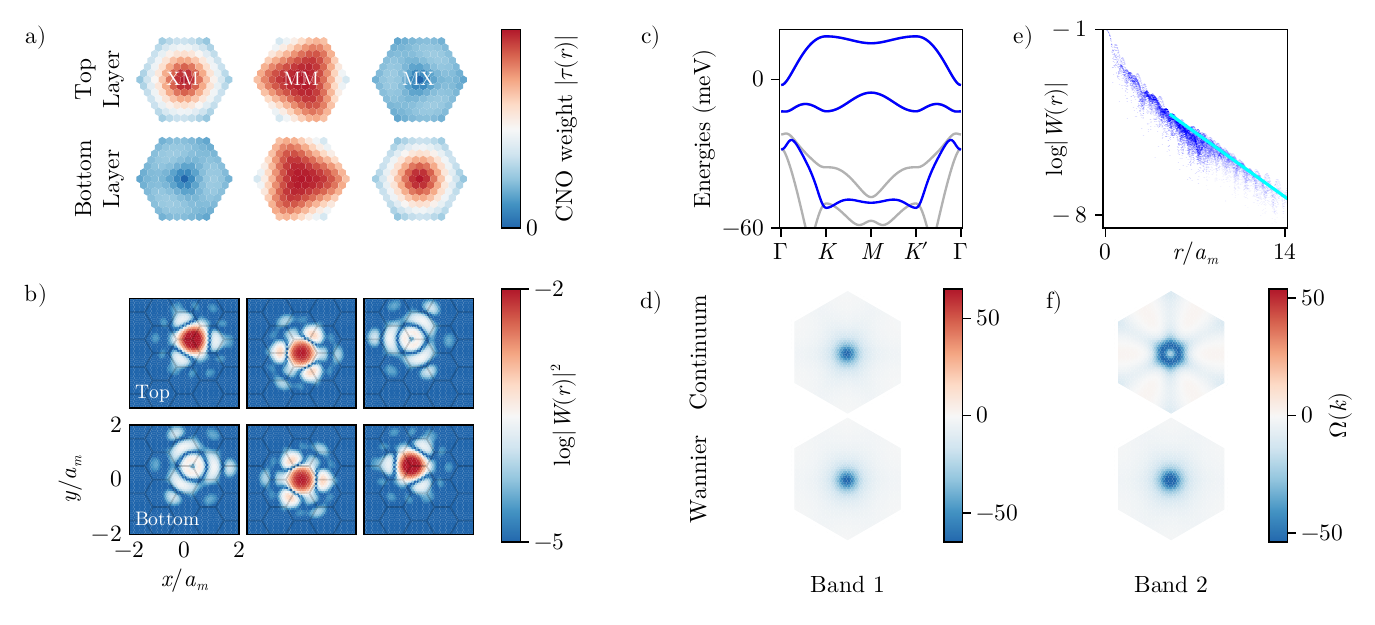}
    \caption{ Local real-space models obtained from the MM-centered CNO subset (top panel) and the multi-centered CNO subset (bottom panel) a) Real-space profiles of CNOs used as trials for Wannierization, plotted separately in the two layers and computed from the top two bands. 
    b) Wannier functions obtained from these CNOs as trial states, plotted in real-space. The Wigner-Seitz {\moire} unit cell is indicated with hexagons. c) Band structure of the 3$^\circ$ tWSe$_2$ Wannier model obtained from the single-shot procedure. This is plotted on top of the band structure of the full continuum model, shown in gray. e) Weight of a representative Wannier function plotted as a function of distance on a logarithmic scale. The Wannier functions decay exponentially, leading to exponentially decaying hoppings in real-space. d), f) Berry curvature of the first and second bands of the full continuum model, compared to the Berry curvature of the first and second bands of the effective 3-orbital Wannier model.}
    \label{fig:wannier_real_space}
\end{figure*}

We first consider the MM-centered partition to construct the UC-1pRDM of the top two bands, shown in Fig.~\ref{fig:twse2_top_two_bands_det_Sk}a).
The top band is dominated by $|\tau_1\rangle$, with a band inversion at $\Gamma$, while the second band has mixed $| \tau_2 \rangle$ and $| \tau_3 \rangle $ character. 
Importantly, the top three CNOs from this partition span the entirety of the two bands.
Notice that individual CNO envelopes have zeros, but the multi-orbital trial set $\{ | \tau^{\textrm{MM}}_1 \rangle, | \tau^{\textrm{MM}}_2 \rangle, | \tau^{\textrm{MM}}_3 \rangle \}$ still spans the two bands as $\det\mathcal{S}_\bk$ (defined in Eq.~\eqref{eq:gram_matrix_S}) is finite as shown in Fig.~\ref{fig:twse2_top_two_bands_det_Sk}b).
\begin{table}[t!]
\centering
\begin{tabular}{l|cccc}
$(C_3, i)$ & $\Gamma$   & $K$ & $K'$ & $M$ \\
\hline
\hline
Band 1 & $(\omega^*,-)$ & $1$           & $1$           & $+$ \\
Band 2 & $(1,+)$        & $\omega$      & $\omega$      & $-$ \\
Band 3 & $(\omega,-)$   & $\omega^*$    & $\omega^*$    & $+$ \\
Band 4 & $(\omega^*,+)$ & $1$           & $1$           & $-$ \\\hline
$\tau_1^{\mathrm{MM}} \equiv s @ \mathrm{MM} (1,+)$ & $(1,+)$ & $1$ & $1$ & $+$ \\
$\tau_2^{\mathrm{MM}} \equiv p_+ @ \mathrm{MM} (\omega,-)$ & $(\omega,-)$ & $\omega$ & $\omega$ & $-$ \\
$\tau_3^{\mathrm{MM}} \equiv p_- @ \mathrm{MM} (\omega^*,-)$ & $(\omega^*,-)$ & $\omega^*$ & $\omega^*$ & $-$ \\
$\left[\tau_1^{\mathrm{MX}}, \tau_1^{\mathrm{XM}}\right] \equiv p_- @ \{\mathrm{MX}, \mathrm{XM}\}$ & $(\omega^*\omega^*, +-)$ & $1\omega$ & $\omega1$ & $+-$ 
\label{tab:CNO_band_characters}
\end{tabular}
\caption{$C_3$ and inversion $i$ characters of the continuum model bands and the CNOs centered on different Wyckoff positions (inversion is a symmetry in the model only where the first harmonic in reciprocal space is considered in the \moire potential and tunneling terms). }
\end{table}

Moving to the MX-centered partition, we see that we would require three pairs of local orbitals to fully span the top two bands, since the top two CNOs lack support on the $\Gamma$-point of the second band (see Fig.~\ref{fig:twse2_top_two_bands_det_Sk}c)).
Recall that this partition breaks $c_{2y}$ symmetry, which manifests in the lack of mirror symmetry in the projections about the $M$-point (the results for XM-centered partition can be inferred immediately by doing this reflection).

We thus have a few choices: a three-orbital model consisting of MM-centered CNOs, a six-orbital model with three each from MX- and XM-centered CNOs, or a mixing between CNOs from different partitions. Here, we consider both the MM-centered CNO subset
\begin{equation}
    {\rm MM \;centered\;set:} \;\{ | \tau^{\textrm{MM}}_1 \rangle, | \tau^{\textrm{MM}}_2 \rangle, | \tau^{\textrm{MM}}_3 \rangle \},\label{eq:MM_centered_CNO_set}
\end{equation}
as well as a multi-centered choice with the dominant CNO from each partition 
\begin{equation}
    {\rm Multi \;centered\;set:} \; \{ | \tau^{\textrm{XM}}_1 \rangle, | \tau^{\textrm{MM}}_1 \rangle, | \tau^{\textrm{MX}}_1 \rangle \}, \label{eq:multi_centered_CNO_set}
\end{equation}
which allows for a complete set of states at all $\bk$, confirmed by the finite $\det\mathcal{S}_\bk$ (Fig.~\ref{fig:twse2_top_two_bands_det_Sk}d)). The total spectral weight of the MM-centered set is $0.755$, while that of the multi-centered CNO subset is $0.899$.
Additionally, while a finite $\det\mathcal{S}_\bk$ guarantees an exponentially localized set of Wannier functions, the degree of the locality is determined by smoothness of $\det\mathcal{S}_\bk$ varies over the Brillouin zone.

We choose to construct two different local models out of the CNO subsets, and carry out a single-shot Wannierization as outlined in Sec.~\ref{subsec:wannier}.
Since the three trial states outnumber the two target bands, the projector $\eta_{\bk m}$ must extend beyond $P$.
We set $\eta_{\bk m} = 1$ for $m \in P$. 
We set $\eta_{\bk m}$ for $m \notin P$ as an exponentially decaying function in band energies.
The extent of the decay is determined by the band inversions outside of $P$, which are revealed in Fig.~\ref{fig:twse2_top_two_bands_det_Sk}a) and c) and equivalently, in the character analysis of the bands at high-symmetry points (see Table~\ref{tab:CNO_band_characters}).
In the multi-centered subset, all three CNOs lack support on the $K / K'$ points in the third band, due to the $C_3$ character mismatch. 
Indeed, the $K/ K'$ point that is closest in energy with the necessary $C_3$ character is in the fourth band, spanned by $|\tau^{\textrm{MM}}_1 \rangle$.
Thus, in the multi-orbital CNO set, $\eta_{\bk m}$ for $m \notin P$ is set to be a decay slow enough to allow for weight from both the third and fourth bands.

Similarly, in the MM-centered subset, the dominant CNO has finite but small support on the $M$-point of the third band, arising from the inversion character mismatch.
If the projector $\eta_{\bk m}$ is taken to be exactly the first three bands, this will lead to an algebraic tail in the Wannier functions. 
Thus, we choose the projector $\eta_{\bk m}$ to also be an exponentially decaying function slow enough to draw weight from the third and fourth bands.

The resulting Wannier functions for multi-centered CNO subset are shown in the bottom panel of Fig.~\ref{fig:wannier_real_space}b). 
They look similar to the trial states, except for tails on the neighboring stacking regions that carry the inter-site hopping.
The Wannier function weights, and subsequently the hopping parameters of the tight-binding model, decay exponentially as shown in the bottom panel of Fig.~\ref{fig:wannier_real_space}e), confirming that the CNO trial states produce a short-ranged model. 
The real-space spreads of the Wannier functions are $0.40, 0.47, 0.40 a_m$ respectively.
The Wannier model produced here is similar to ones in previous works~\cite{Crepel2024_prr, Tuo2025_theory_topological_superconductivity}, but is informed by the intrinsic character of the target bands and the EBRs induced by local orbitals located at Wyckoff positions. 
Reproducing the dispersion is a weaker test than reproducing the wavefunctions, so we also compare the quantum geometry of the truncated model against the continuum result.
Fig.~\ref{fig:wannier_real_space}d) and e) (bottom panel) show the Berry curvature $\Omega(\bk)$ of the two lowest bands of the tight-binding model and of the continuum model over the {\moire} Brillouin zone.
The Berry curvature of the first band is reproduced both qualitatively and quantitatively well, while the second band deviates slightly in the qualitative profile.

The corresponding figure for the MM-centered CNO set is shown in the top panel of Fig.~\ref{fig:wannier_real_space}.
Due to the $C_3$ character of $| \tau^\textrm{MM}_2 \rangle$ and $| \tau^\textrm{MM}_3 \rangle$, the Wannier functions obtained from these two CNOs are forced to have a zero at their center.
Compared to the multi-centered set, the Wannier functions are slightly more spread out in real-space but are exponentially localized to almost the same degree, seen in the top panel of Fig.~\ref{fig:wannier_real_space}e).
The real-space spreads of the Wannier functions are $0.44, 0.56, 0.53 a_m$ respectively.
The Berry curvature computed from this local model is qualitatively similar to the multi-centered model.

CNOs offer a systematic way to construct local tight-binding models. 
The choice between the two models depends on the numerical method used to treat interactions.
The multi-centered model has more localized Wannier functions and therefore yields more local interaction parameters. 
However, because the three orbitals are centered at different sites, a cluster DMFT ansatz that respects $C_3$ symmetry would require a seven-site cluster.
In contrast, the MM-centered model requires only one site with three orbitals, making it easier to preserve the $C_3$ symmetry.
The trade-off, of course, is that the MM-centered model has longer-range interaction terms.

\section{Conclusion}
We have used the unit-cell one-particle reduced density matrix to extract local orbitals from a band manifold.  
Conventionally, such orbitals are inferred from the local quantum chemistry of the material or from EBRs based on symmetry eigenvalues of the wavefunction at the high symmetry points. 
The CNO construction reverses this procedure and gives orbitals ranked by their spectral weight.
In this sense, CNOs provide a direct measure of the multi-orbital character for a set of target bands.

The eigenvalues, $\{ \lambda_a \}$, simultaneously capture the singular values of the map between the local unit-cell and target band subspace, and the entanglement spectrum associated with a unit-cell partition. 
The CNO envelope functions, $\{ s_{a, n, \bk} \}$, resolve how the local orbitals are distributed across momentum space.
The zeros in envelope function arise from obstructions to a local description. 
While topology enforces irremovable zeros, symmetry-forced zeros are associated with an incompatible choice of real-space center. 
The latter can be removed by moving the partition to the appropriate Wyckoff position.

A symmetry-preserving partition assigns each CNO a Wyckoff position and a site-symmetry representation, interpolating between local orbitals selected by the wavefunctions and the EBRs of topological quantum chemistry. 
The resulting overlap criterion then tests whether these local degrees of freedom span the target bands throughout the Brillouin zone. This combines symmetry compatibility with the stronger requirement of reproducing the full wavefunction structure, that is, the quantum geometry.

Our construction can be applied to both lattice and continuum models. 
In the BHZ model, the appearance of an irremovable zero in the CNO envelope function gives a real-space manifestation of the Wannier obstruction.
In monolayer WSe$_2$, the zeros reveal the obstructed atomic character of the bands. 
For twisted bilayer WSe$_2$, the CNOs identify a set of local degrees of freedom associated with the MM, MX, and XM stacking regions. Their occupations and momentum-space weights evolve continuously with twist angle and reorganize when the topology of the low-energy bands changes. The same CNOs provide compact trial states for Wannierization, yielding a short-ranged model that reproduces not only the dispersion but also the Berry curvature of the continuum bands.

Beyond the single-particle model, the CNOs provide a basis for the interacting
problem. Because they diagonalize the unit-cell correlation matrix, they are the
natural basis for band-projected interactions \cite{verma2025local}. 
The most localized CNO carries
the largest on-site interaction, so the occupation hierarchy orders the
interaction scales and gives a wavefunction-derived route to multi-orbital Hubbard models.

The same ranking addresses the choice of local orbital in quantum embedding.
Methods such as dynamical mean-field theory \cite{RozenbergGeorges1996} and density-matrix embedding \cite{Knizia2012} require a correlated subspace onto which the many-body problem is projected \cite{Karp2021DependenceOrbitals}. 
The occupation spectrum instead ranks local orbitals by their weight in the manifold, so the leading CNO defines a impurity space fixed by the wavefunctions. Their locality makes them natural candidates for the ansatz entering the dynamical self-energy.

Because the only input is a projector, the construction applies without modification to wavefunctions from first-principles calculations, where the occupation spectrum could accompany the band structure as a routine measure of multi-orbital character. 
The leading CNOs also provide variational building blocks for the competition between fractional Chern insulator and charge-density-wave order, which has recently been observed in rhombohedral graphene \cite{Aronson2025, Butler20261, Lu2024FractionalGraphene}. We anticipate that the dominant orbital favors charge-density-wave order, while the subdominant modes stabilize a fractional state. 
A precise calculation is needed to establish this connection, which we leave for future work.

In sum, our construction shows that the choice of local orbitals need not be imposed before analyzing a band manifold. 
It can instead be inferred from the manifold itself, while respecting locality, symmetry, quantum geometry, and topology. 
We expect this perspective to be useful for connecting first-principles wavefunctions and continuum models to the local degrees of freedom used in minimal models of correlated and topological quantum materials.

\section*{Acknowledgments}
We gratefully acknowledge discussions with Lukas Muechler, Carolina Paiva, Ammon Fischer, Julian Ingham, and Nicolás Morales-Durán.
This material was supported by the National Science Foundation (NSF) under CAREER Award No.~DMR-2340394, the Materials Research Science and Engineering Centers
(MRSEC) program through Columbia University under the
Precision-Assembled Quantum Materials (PAQM) Grant No. DMR-2011738; Sloan Research Fellowship (FG-2025-24714), and the Army Research Office under Grant No.~W911NF-26-1-A283. The Flatiron Institute is a division of the Simons Foundation.

\bibliography{CNO_construction}

\end{document}